# Study of Cu₂O Particle Morphology on Microwave Field Enhancement

T. D. Musho,[a,b,d,*] C. Wildfire, [a,d] N. M. Houlihan,[c] and E. M. Sabolsky, [b,d] D. Shekhawat [a]

The dielectric enhancement and modulation of a cuprous oxide (Cu2O) microwave-active catalyst material is investigated from an experimental and computational point of view. Experimental synthesis of two particle morphologies that included a cube and spike were carried out using an EDTA hydrothermal synthesis method. The permittivity for the spiked particles at low volume fraction in a paraffin composite exhibited a 20% increase when compared to the cube-shaped particles at the same volume fraction. Using a finite difference time domain (FDTD) simulation, the improvement in permittivity was attributed to the enhanced electric field near the tip of the spike particles and the neighboring interaction at higher volume fractions. The increased electric field at the tips of the particles induces a change in polarizability (dipole density) within the matrix material, which increases the effective dielectric properties of the composite. Furthermore, it was determined that an electrically conductive particle within a high permittivity matrix material is advantageous for generating high localized electric fields that can be utilized for microwave-assisted catalytic reactions.

## Introduction

Over the past decades, there has been an increased number of studies related to microwave-assisted chemical processes[1,2,3]. A market that has seen early adoption of this approach is the conversion of chemical feedstocks to value-added chemicals. More specifically, the application of microwave processes to aid the pyrolysis (thermal cracking) and gasification of biomass[4]. Looking at challenging gasification process, the most common method of converting feedstocks to other products is through a traditional thermal approach typically at an extremely high temperature and pressure[5,6]. In most scenarios, the reaction is not spontaneous and requires the input of energy to evolve the reaction and this energy is typically provided in the form of thermal energy. However, the thermal conversion process requires special reactors that can handle high pressure and temperature streams, with the additional burden of energy required to achieve these pressures. In providing an alternative approach, the following study is focused on a microwave-assisted catalytic approach that is aimed at generating extremely highly localized electric fields on the catalyst that permit new processing windows. These localized fields emulate the high potentials not achievable from a traditional catalyst and the new processing window permit the streams to be at lower temperature and pressures. This, in turn, provides a scenario where the microwave-assisted catalytic approach (see Figure 1a for illustrative setup) could outperform other gasification approaches.

The microwave assisted approach targeted in this study is to use a fixed (single) wavelength microwave source to excite a tailored dielectric catalyst surface. The approach will be to understand how the dielectric properties can be tailored as a function particle geometry for a given material composition and how the field interacts with particular surface features. The primary hypothesis is that the particle morphology can be used to achieve dielectric properties greater than the simple mixing of the dielectric properties of the constitutive material within the composite. The secondary hypothesis is that the sharp corners of the particles will provide field enhancement to increase the electric field an in turn achieve enhancement of the process.

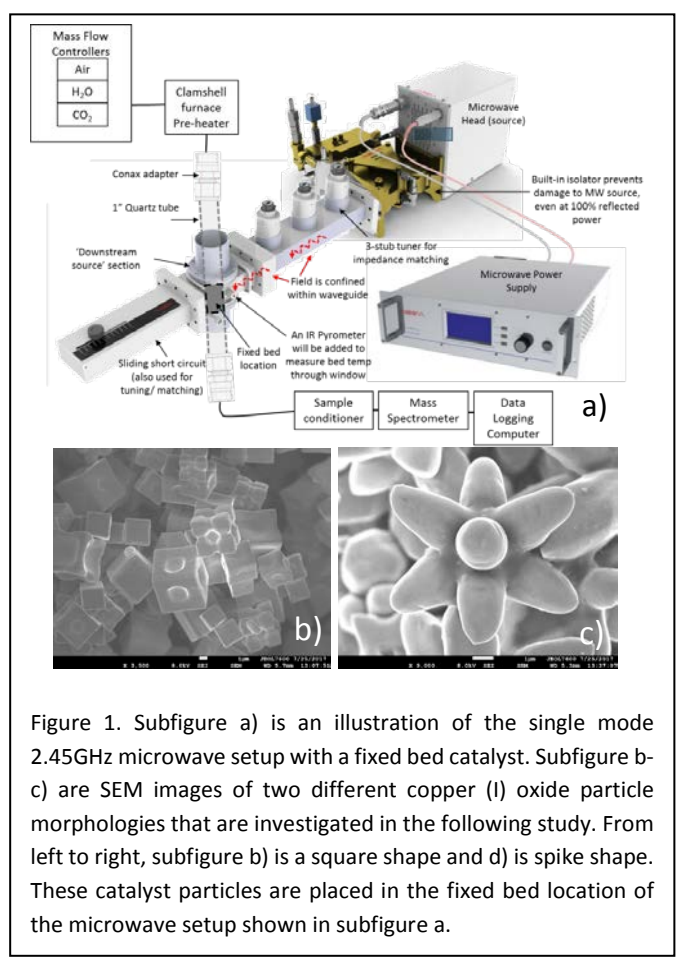

Figure 1. Subfigure a) is an illustration of the single mode 2.45GHz microwave setup with a fixed bed catalyst. Subfigure b-c) are SEM images of two different copper (I) oxide particle morphologies that are investigated in the following study. From left to right, subfigure b) is a square shape and d) is spike shape. These catalyst particles are placed in the fixed bed location of the microwave setup shown in subfigure a.

[a.] National Energy Technology Laboratory, Morgantown, WV 26507, USA.
[b.] West Virginia University, Materials Science and Engineering Department, Morgantown, WV 26506, USA.
[c.] SUNY Polytechnic Institute College of Nanoscale Engineering and Technology Innovation, Albany, NY 12203, USA.
[d.] Oak Ridge Institute of Science and Education, Oak Ridge, TN 37830, USA.
* Corresponding Author: tdmusho@mail.wvu.edu, P.O. Box 6106 Morgantown, WV 26506-6106, USA.

**Previous Particle Studies**

There has been a reasonable number of studies that investigated the effects of particle shape on microwave absorption within a composite material[7,8,9,10]. Liu et al. demonstrated enhanced microwave absorption performance using monodispersed triangular iron oxide nanocrystals versus spherical or cubic particles due to interface polarization between the particles[7]. They went as far to experimentally demonstrate that a peak loss is achievable at frequencies of approximately 12GHz. It was interesting to find that their iron oxide nanocrystals were fairly small at 8nm and they were able to achieve resonance. Ma et al. conducted a similar study with nickel nanoparticles[8]. They found that there was no resonance for nanospheres but a strong resonance at 8GHz associated with conical nanorods. This was an interesting finding that provides justification for spiked shaped particles. The size of their particles was larger than the iron-based particles with a length of the rod of 0.5μm and the diameter of approximately 50nm. However, they did not conduct the study at different volume fractions, which makes it difficult to discern the neighboring interactions from their data. Another study conducted by Myllymaki et al. investigated attempts to provide information about neighboring interaction of sphere-, flake-, and disc-shaped $BaFe_{12}O_{19}$. Their particles were less than 2μm. They did not achieve a discernible resonance at 7GHz but did demonstration show some variation in the shape. They also demonstrated that the volume fraction influences the permeability and loss tangents. A common theme across many of these studies is their materials are magnetic. This means their material will interact with both incident electric and magnetic field components. It is interesting to point out that their permeability is not huge and they are still able to achieve resonance, granted the frequencies they achieved resonance is much greater than 2.45GHz. As a result, it is reasoned that our particles will have to be larger to interact with a larger wavelength of incident electromagnetic radiation.

These previous studies provide a starting point for further investigations. Hence, the proposed study is interested in answering why and how the volume fraction influences the properties. We have selected a material with a magnetic response, which is similar to previous studies. We will rely on achieving strong coupling of the electric field. A detailed description of material properties can be found in the following sections.

**Microwave Based Approach**

As discussed in the introduction, it is important to point out the potential key advantages of a microwave-based approach. Microwave-based chemistry in the form of pyrolysis has shown good success in the conversion of biofuels with significant rate enhancement[2]. The rate enhancement realized in these pyrolysis experiments is attributed to thermal/kinetic effect. Typically, in these experiments, the advantage of the microwave comes in the ability to superheat the solvents at atmospheric pressures, selectively heat heterogeneous catalysts, and elimination of wall effects cause by an inverted temperature gradient[11]. With these advantages comes a new processing window that allows reaction with different product distribution, cleaner reaction profiles, and higher yields. Beyond the thermal enhancements, there are significant advantages in tailoring the catalyst to produce extremely high localized electric fields in heterogeneous catalysts. These fields aid in deformation of the molecules on the catalyst surface and repealing products from the surface to increase the kinetics of the reaction.

**Electromagnetic Theory and Interactions**

Understanding the dielectric properties of a heterogeneous catalyst is critical to understanding the response of catalyst during irradiation. In the medium-frequency regime ($10^9$-$10^{12}$Hz) dipole effects and bond relaxation control the permittivity and permeability of a material. Focusing on the dipole effect, an important quantity is the polarizability of a material. In linear materials, the electric polarizability takes the following form,

$$\boldsymbol{P} = \chi_e \varepsilon_0 \boldsymbol{E} = (1 - \varepsilon_r)\varepsilon_0 \boldsymbol{E}, \quad (1)$$

where P is the polarization vector, χ is the electric susceptibility, $\varepsilon_r$ is the complex permittivity ($\varepsilon_r$= ε'+ iε"), and E is the electric field. The polarization vector can be interpreted as the volume density of electric dipoles within the material. Likewise, on the magnetic side, the magnetic polarizability takes the following form,

$$\boldsymbol{M} = \chi_m \mu_0 \boldsymbol{H} = (1 - \mu_r)\mu_0 \boldsymbol{H}, \quad (2)$$

here M is the magnetization vector, H is the magnetic field, $\chi_m$ is the magnetic susceptibility and $\mu_r$ is the complex permeability ($\mu_r$=μ'+iμ"). Following the constitutive relations, when an electromagnetic field incidents a material the electric flux density is proportional to the applied electric field plus the induced polarizability ($\boldsymbol{D} = \varepsilon_0 \boldsymbol{E} + \boldsymbol{P}$). Similarly, the magnetic flux density follows a similar relation ($\boldsymbol{B} = \mu_0 \boldsymbol{H} + \boldsymbol{M}$). Therefore, the objective of this study is to tailor the polarizations such that they are in phase with the incident fields to maximize the electric flux density, which will influence the catalysis process.

The microwave setup that is currently being used is a single mode microwave setup with a frequency centered around 2.45GHz. Figure 1 is an illustration of the test section of the microwave. The $Cu_2O$ particles are placed within a quartz tube in the fixed bed location and the gas is flowed down over the particles while being bombarded by microwave radiation. One application that this approach is currently targeting is to convert hydrogen (syngas) to longer chain hydrocarbons. While it is beyond the scope of this research to elaborate on the chemistry details of this process, the focus of this research is to understand the fundamental microwave interaction with the catalyst particles. We employ a combination of experimental and computational techniques to study this interaction.

## Materials Selection

For this study, a copper (I) oxide (Cu₂O) or cuprous oxide was synthesized with two different shape morphologies. These morphologies include cubed and spike. Figure 1b,c are SEM images of experimentally synthesized particles. Cuprous oxide exhibits diamagnetic properties (expels magnetic fields) at low temperatures (<5K) and for octahedron shapes[12]. While this might be an advantageous property, at room temperature all the cuprous oxide shapes exhibit ferromagnetic properties. The ferromagnetic properties arise from the spin polarization or difference in spin population of the spin-up and spin-down electrons. The unit cell of cuprous oxide is cubic with oxygen situated on a body-centered lattice and copper on a face-centered sub-lattice. It is also important to point out that cuprous oxide is a p-type semiconductor that exhibits moderate electrical conductivity that is mostly attributed to small polaron hopping[13] and the partial pressure of oxygen. The material exhibits a moderate direct band gap of 2eV[14].

## Experimental and Computational Details

### Particle Synthesis

Various methods of synthesizing several morphologies of Cu₂O of well-defined shapes have been reported in several publications[15,16,17]. Attempts to reproduce the results were varied but generally, the methodology provided by Xu et al. were followed[18]. All chemicals were of analytical grade and used without further purification. The molar ratios of Cu(I)-EDTA were calculated Copper (ll) Nitrate (Sigma-Aldrich, 98%) and Ethylenediaminetetraacetic Acid, trisodium salt (Acros Organics, 98%) were dissolved into 80 ml of DI water. The pH of the solution was adjusted by adding NaOH (Alfa Aesar, 97%). The solution was then transferred to a Teflon-lined stainless steel autoclave and heated to the temperature and specific dwell time. The particles were then washed with water and absolute ethanol, respectively and dried at 50°C.

The phase of the transformed powder was characterized using a PanAnalytical X'pert Pro X-Ray diffraction system (PW 3040 Pro) using a wavelength of Kα 1.54184 Å. Scans were performed with a 0.026°/s scan rate and a 120 s step time. XRD patterns were analyzed using X'Pert Highscore Plus which is based on the Rietveld refinement method using a pseudo-Voight function to model the peak profile. Characterization of the particle size and morphology were performed using a scanning electron microscope (SEM, Hitachi S-4700).

### Experimental Dielectric Measurement

Dielectric measurements were made with a 7 mm diameter coaxial airline (HP model no. 85051-60010) connected to a Keysight N5231A PNA-L microwave network analyzer. The composites samples were prepared by uniformly mixing the Cu₂O powders in paraffin with a 1, 5, 10, and 15 vol% Cu₂O. The mixture was formed into a cylindrical plug with an outer diameter of 7mm, inner diameter of 3 mm and a thickness of 18 mm. The permeability and permittivity of the composites were measured in the range of 1-5 GHz. The frequency range was restricted to 1-5 GHz to avoid resonance issues within the coaxial test cell. There was not significant variation over this frequency for the given material and therefore on the 2.45 GHz frequency data will be reported.

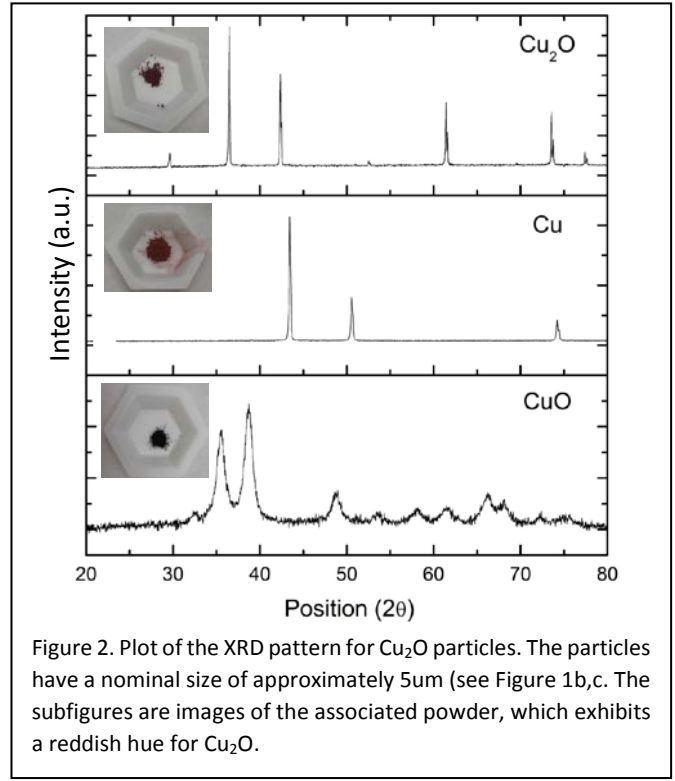

Figure 2. Plot of the XRD pattern for Cu₂O particles. The particles have a nominal size of approximately 5um (see Figure 1b,c. The subfigures are images of the associated powder, which exhibits a reddish hue for Cu₂O.

### Computational Model

A finite difference time domain (FDTD) was used to predict the polarizability of the particles and the scattering parameters of a particle/paraffin composite material. COMSOL RF Multiphysics commercial software was employed to conduct these studies. Maxwell's equations were solved in the frequency domain. By using the frequency domain approach the computational expense is decreased compared to the transient domain as a result of a complex value representation and the sinusoidal signal assumption that can be represented using Fourier analysis. Under the frequency domain assumption, we are forced to assume all dielectric properties are linear (nonlinearities would be the change in dielectric properties with the intensity of electromagnetic field). The equilibrium frequency domain equation that is solved take the following form,

$$\nabla \times \frac{1}{\mu_r}(\nabla \times E) - k_0^2 \left(\varepsilon_r - \frac{i\sigma}{\omega\varepsilon_0}\right) E = 0. \quad (3)$$

| Particle Geometry | ε' | ε" | µ' | µ" | tan(δ_ε),σ=0 | tan(δ_µ),σ=0 |
|---|---|---|---|---|---|---|
| Paraffin 100vol% | 2.2023 | 0.0006 | 0.9934 | 0.0048 | 0.0003 | 0.0048 |
| Cubes 5vol% | 2.2813 | 0.0148 | 0.9980 | 0.0085 | 0.0065 | 0.0085 |
| Spikes 1vol% | 2.2577 | 0.0119 | 0.9971 | 0.0056 | 0.0053 | 0.0056 |
| Spikes 5vol% | 2.2903 | 0.0179 | 1.0075 | 0.0096 | 0.0078 | 0.0095 |
| Spikes 10vol% | 2.3556 | 0.0160 | 1.0064 | 0.0104 | 0.0068 | 0.0103 |
| Spikes 15vol% | 2.4283 | 0.0167 | 1.0126 | 0.0079 | 0.0069 | 0.0078 |

Table 1: Experimentally determined dielectric properties of paraffin and $Cu_2O$ particles mixture at 2.45GHz and room temperature. These values were experimentally determined using a network analyser with coaxial dielectric measurement technique and results were averaged over sample lengths. The trend demonstrated in the table is with increases complexity shape complexity (cube->spike) the effective permittivity and permeability increase. For the spike particles, as the volume fraction increases the effective dielectric properties increase greater than what is predicted from simple mixing relationships. This is determined to be a result of neighbouring interactions that increase the dipole density (polarizability) between particles and spike tips, see Figure 3.

Here $k_0$ is the wave number of free space, σ is the electrical conductivity, and ω is the angular frequency. The term containing the electrical conductivity defines the finite conductivity loss, dipole losses are contained within the imaginary terms.

The domain was constructed using a single unit cell representation of the particle surrounded by paraffin matrix. Reflective boundaries on the lateral edges and open boundaries on the top and bottom (ports). The frequency was specified to be 2.45GHz and the fields and scattering parameters were monitored at this frequency.

The material properties are important when modeling the response of a composite material. In this study, the complex dielectric properties, which include the real and imaginary parts of both the permittivity and permeability were provided along with the conductivity. For $Cu_2O$ those properties were ε=8.8000+0.0821i and µ=1.0000+0.0240i. For paraffin, those properties were ε=2.2023+0.0006i and µ=0.9934+0.0048i. The $Cu_2O$ properties were determined based on literature values and the values for paraffin were based on the experimentally determined values, see Table 1. It is important to point out that the permittivity of the particle is greater than the permittivity of the paraffin. However, the particle has more loss than the paraffin. An additional aspect that is often overlooked is the conductivity of each of the materials. We assumed an electrical conductivity of 20 S/m for cuprous oxide and a conductivity of nearly zero at $1×10^{-13}$ S/m for paraffin. This is an important aspect because a current will be generated within the conductive particle that is proportional to the conductivity times the electric field.

The effective dielectric properties of the composite were calculated from the scattering parameters using a similar method that is employed by the experimental network analyzers. The Nicholson-Ross-Weir method[19] was used that provided a direct calculation of the permittivity and permeability from the scattering parameters.

## Results and Discussion

The XRD pattern of the dried particles in Figure 2 illustrates the cubic $Cu_2O$ (JCPDS File 78-2076) without any noticeable impurities in the phase[14,20]. The morphologies of the particles were further analysed by SEM as shown in Figure 1b,c. Each batch of particles had similar particle size ranging from 5-10µm and had good uniformity in particle morphology. During the synthesis process, EDTA acts as both a reductant and chelating agent. The morphology is closely dependent on the Cu:EDTA ratio, pH, temperature, dwell time, and pressure. The cubic particles in Figure 1b range from 5-10µm and were a mixture of solid cubes and with a shallow void space on the centre faces of the cube due to the rapid growth of the outer crystal planes[12]. The second morphology synthesized is a faceted 8-pod cubic structure that grew to around 5 µm and have more angular features. The third morphology synthesized was a spiked particle as shown in Figure 1c. The arms of the particle are slightly rounded but provide a wide angle between the spokes and sharp point at ends.

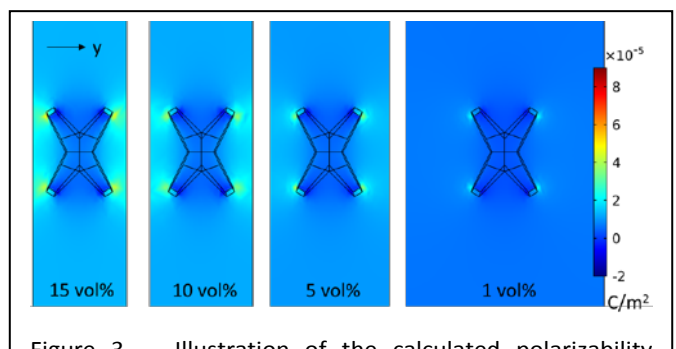

Figure 3. – Illustration of the calculated polarizability (dipole density) in the y-direction for a spiked $Cu_2O$ particle at four difference volume fractions. As the volume fraction increase the density of dipoles within the paraffin increase resulting in an effective increase in dielectric properties. This is direct result of the tip geometry and the interaction of neighbouring particles.

In order to understand the effect shape has on the electromagnetic field, the particles were set in paraffin at low volume percentages, well below the percolation limit to avoid the particles forming a conductive network. If the particles are able to form a conductive network the simple mixing dielectric properties are recovered. Table 1 is a summary of permittivity and permeability of each morphology taken at 2.45 GHz and room temperature. As mentioned in the previous section, at room temperature $Cu_2O$ is a weak ferromagnetic and does not exhibit significant coupling with the associated magnetic field at this frequency. This is quantified with permeability values near unity. However, the permittivity values exhibit a noticeable modulation as a function of volume fraction and shape.

While Table 1 provides an overview of the experimentally determined values of the permittivity, the significance of the influence shape on the effective properties is not obvious. By taking the effective paraffin/particle properties and the known volume fraction of particles within the mixture, the dielectric properties of the particles can be calculated. This is carried out by solving the Maxwell-Garret effective medium mixing equation[21] for the dielectric properties of the particle. Typically, this mixing equation is only valid for low volume fractions of mixtures, which is an additional reason why the volume percentage is kept below 15% for this study. In carrying out this analysis, the dielectric properties for a cube shape is calculated to be ε=3.7891 at 5 vol% and ε=4.5586 for the spike at 5 vol%. This equates to nearly a 20% increase in permittivity by changing the shape of the particle from a cube to a spike. This is a significant finding in that it provides evidence that the dielectric properties can not only be controlled by bulk material selections but also by controlling the particle morphology. While this similar finding has been demonstrated for other material at other frequencies, this is the first demonstration for $Cu_2O$ at microwave wavelengths.

The reasoning for the significant increase in permittivity values is due to localization of the fields of the material. More specifically, the increase in electric fields with near the tips of the spike particles. These localized electric fields increase the dipole density or polarizability within the paraffin in the proximity of the particle tips. Referring to Equation 1 for a given electric field, an increase in polarizability must be directly related to an increase in complex dielectric constant. To add an additional aspect, the distance between particles, which is proportional to the volume fraction also influences the ability to generate electric fields between particles. This can be seen in Table 1 for the spike particles at difference volume fractions. At higher volume fractions the spacing between the particles is decreased and this increases the electric field within the paraffin. This increase in the electric field between particles manifests itself with an increase in measured effective dielectric properties. Further justification for these claims can be made by modeling the fields as will be discussed in the next section.

**Simulations of Microwave Interaction with Particles**

As detailed in the previous section, an FDTD method was employed to investigate the fields and predict the effective dielectric properties as a function of particle geometry and volume fraction. A useful quantity that is described by Equation 1 above is the polarizability. The polarizability is described as dipole density within the material. The more dipoles that are able to be formed per unit area is governed by the dielectric properties of the material and the field strength.

Figure 3 is an illustration of the polarization in the y-direction for a 5μm spike particle at four different volume fractions associated with the experimental values provided in Table 1. The color bar associated in Figure 3 represents the polarization density in a plane that is placed at the intersection of the spike tips. These contour plots are associated with a given snapshot in time or phase when the polarization is maximum. It can be envisioned that these dipoles are turned on and off as the incident field (at 2.45GHz) interacts with the particles inducing a dipole in the paraffin. This is not intuitive in this case because the particles have a larger permittivity than the paraffin. Therefore, the particles are more capacitive than the paraffin but as illustrated in Figure 3 the dipoles are generated within the paraffin. This is a result of the particles having such a high conductivity that they charge polarize quickly generating a current within the particle. This can be confirmed by plotting the current density from the simulation. If the conductivity of the particle was not as large as it is, the polarization would actually be greatest within the particles, resulting in a different response. It is advantageous to have a conductive particle that quickly polarizes and can essentially act as a perfect electrical conductor. In a perfect electrical conductor, the surface charge must be proportional to the charge within the solid. This maximizes the electric field around the particle.

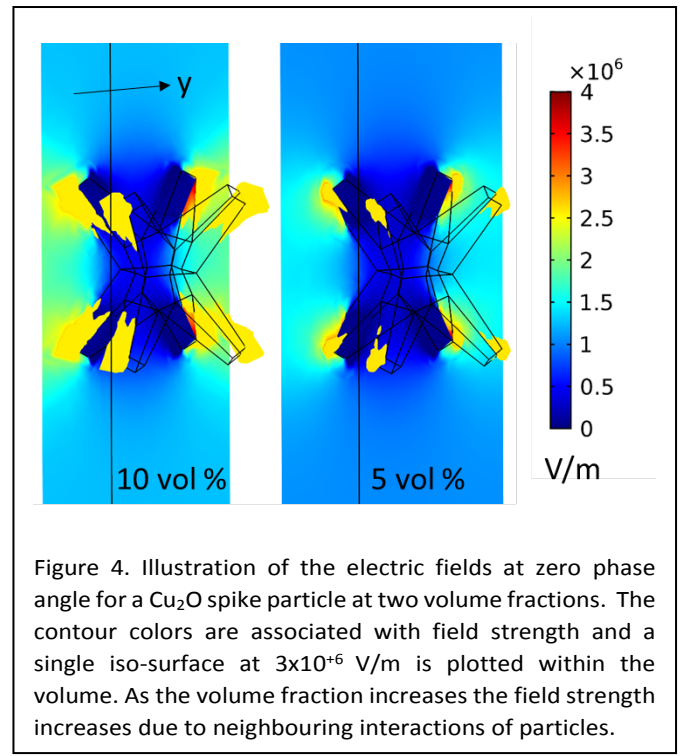

Figure 4. Illustration of the electric fields at zero phase angle for a $Cu_2O$ spike particle at two volume fractions. The contour colors are associated with field strength and a single iso-surface at $3\times10^{+6}$ V/m is plotted within the volume. As the volume fraction increases the field strength increases due to neighbouring interactions of particles.

Referring back to Figure 3, in addition to the increase dipole density between particles with increasing volume fraction (decrease spacing), it is important to highlight the location of the maximum. It is noted that the high dipole density is within the paraffin near the spike tips.

There are arguably two influences that have taken place that increase the effective dielectric properties. Those two influences are the particle shape and the proximity of the particles relative to each other (volume fraction). Figure 3 illustrates both of these aspects. Focusing on the lowest volume fraction spike particles, 1% volume fraction, there is increased polarization near the tip of the particle in the paraffin. Looking across the subfigures there is a drop in the polarization with decreasing volume fraction or increasing spacing.

Another means of quantifying the tip enhancement is to plot the electric field. Figure 4 is a plot of the electric field for two volume fractions. In this figure, the electric field is plotted on a plane that intersects the tips and iso-surfaces (surface of the constant field, $3 \times 10^{+6}$ V/m) are plotted. Here the highest associated electric field within this region is focused near the tips, similar to the polarization. This confirms the previous claim that a localized field enhancement is realized at 2.45GHz and it is located at the tips. The second influence mentioned in the previous paragraph is also demonstrated in electric field plot as an increased field between particles. The electric field within the particle is low because of the high conductivity of the particle.

By varying the material properties within the model the field and scattering parameters can be studied. The results of increasing effective dielectric properties that was experimentally determined and presented in Table 1 can be reproduced by taking the scattering parameter from the simulation and apply Nicholson-Ross-Weir method. However, in order to achieve a 10% increase in the effective properties and 20% increase in particle dielectric properties all possible orientations of the particles must be accounted. As shown illustrated Figure 3 and 4, only one orientation was provided in this analysis. The highest electric field is actually achieved when the tips of the particle are oriented at each other. A follow-on study will consider all possible orientation and the associated effective properties.

## Conclusions

In this study, the influence of the shape on the dielectric properties was studied at 2.45GHz. The proposed application of these particles is to be a single mode microwave catalyst for gasification purpose. Two morphologies were synthesized that included cube and spiked. These were mixed with paraffin and the dielectric properties were experimentally determined for several volume fractions of particles. Based on the experimental findings, it was determined that dielectric properties could be modulated up to 20% through shape and volume fraction selection. This enhancement was reasoned to be a result of localized fields generated near the particle tips and between particles. This was confirmed through finite difference simulation of a single particle. It was confirmed that the increase in dielectric properties was a result of both localized fields near the tip and the increase in the field between particles. Further study using the computational model confirmed that having an electrically conductive particle with field enhancement geometries is best for generating high fields between particles within the dielectric matrix material. This reasoning holds true when the dielectric properties of the particle are greater than the matrix material.

## Conflicts of interest

There are no conflicts to declare.

## Acknowledgements

This research was supported in part by an appointment to the National Energy Technology Laboratory Research Participation Program, sponsored by the U.S. Department of Energy and administered by the Oak Ridge Institute for Science and Education. The authors also would like to acknowledge Amy Falcon and Allison Arnold for their assistance.